\documentclass[twocolumn,preprintnumbers,amsmath,amssymb]{revtex4}
\usepackage[]{graphicx}

\usepackage{amsmath}
\usepackage{amssymb}
\usepackage{amsthm}
\usepackage{amsfonts}
\usepackage{enumerate}
\usepackage[colorlinks=false]{hyperref}
\usepackage{centernot}
\usepackage{tikz}

\begin{document}

\title{Anomalous magnetic transport and extra quantum oscillation in semi-metallic photon-like fermion gas}

\author{Xi Luo$^1$}
\thanks{These two authors contribute equally.}
\author{ Fei-Ye Li$^{2,3,4}$}
\thanks{These two authors contribute equally.}
\author{ Yue Yu$^{2,3,4}$}
\affiliation {1. College of Science, University of Shanghai for Science and Technology, Shanghai 200093, China\\
2.Center for Field
Theory and Particle Physics, Department of Physics, Fudan University, Shanghai 200433,
China \\
3. State Key Laboratory of Surface Physics, Fudan University, Shanghai 200433,
China\\
4. Collaborative Innovation Center of Advanced Microstructures, Nanjing 210093, China}

\begin{abstract}
{In the absence of Lorentz symmetry, the pseudospin-1 counterpart of Weyl fermion (feroton) with linear dispersions and an exact flat band can emerge in condensed matter systems. The flat band branch of feroton is equivalent to the longitudinal photon in Maxwell theory, which is a redundant degree of freedom due to the emergent (fermionic) gauge symmetry. Upon coupling to an external magnetic field, the fermionic guage symmetry is broken and the flat band ferotons become gapless excitations characterized by Landau level indices ($n>1$). In the long wave length limit, these gapless modes are of the opposite chirality to the chiral anomaly related zero Landau level, which leads to much more plentiful magnetic transport properties. To further explore the novel properties of these gapless modes, we investigate the quantum oscillation through a generalized Lieb lattice model. We find an extra oscillating behavior which indicates the existence of these exotic gapless modes.} We collect known {\it ab initio} calculation data from literature and {discuss} the possibility of {realizing} the semi-metallic feroton gas in real materials.
\end{abstract}

\date{\today}

\maketitle

\section{Introduction }

Spin-statistics connection for elementary particles which constitute our universe is a general relation based on the Lorentz symmetry of space-time \cite{ssc,ssc1}, {while this relation is not necessary for elementary excitations in condensed matter because of the absence of Lorentz symmetry.} As a counterpart of Weyl semimetals (WSMs) \cite{wan}, the pseudospin one semimetal was proposed \cite{2dspin1,LY1} and predicted in the crystals with the symmetry of space groups 199, 214 and 220 \cite{ber}.  The spectrum of pseudospin one fermions is the same as that of photon and so we dub them the photon-like fermions, shorten as  {\it ferotons}.  The many feroton system forms a semi-metallic gas and is named the feroton semimetal here. Experimentally, its doubled version with genuine spin, the massless Kane fermions with {a sixfold degenerate point},  was observed  \cite{nphy,ncomm}.  The interesting magnet-optics of massless Kane fermions was studied \cite{magop,magop1}.  In a topological semimetal MoP, another kind of triply degenerate fermion was observed  \cite{nature}. Subsequent studies searching for the material realizations of triply degenerate fermions and their novel properties were done \cite{BG,tdnp1,tdnp2,tdnp3,tdnp4,tdnp5,tdnp6,tdnp7,tdnp8}.  Among these triply degenerate fermion systems, the  ferotons are of particular interest as we will see.

Comparing with the spectrum of  WSM, there is an {exact} flat band besides two linear bands for the feroton semimetal. {The emergence of flat band usually implies new physics. Especially, the researcher's} enthusiasm on the flat band is stimulated due to its importance in the topological states of matter \cite{fci4,fci5,fci6}.  The fractional quantum anomalous Hall effects were predicted  in nearly flat bands \cite{fci1,fci2,fci3}. The materials supporting a topological non-trivial flat band  have been designed for organometals \cite{orgm,orgm2}.

{The exact flat bands also emerge in several two-dimensional tight-binding models which are based on Lieb lattice model  \cite{Lieb,aoki,Deng,Miy,WuCJ,Berc}.} They are designed and  experimentally realized in cold atoms on optical lattices. A photonic crystal Lieb lattice has been experimentally demonstrated \cite{photonic1,photonic2}. {The flat band in a 2D Lieb lattice} may also be simulated by circuit quantum  electrodynamics simulator \cite{WuYin}.
 Because of the infinite degeneracy of the flat band and  the singular density of states (DOS) at Fermi surface, many astonishing new phenomena may emerge. Wigner crystals \cite{WuCJ}, superfluids \cite{volovik, Zhai,Julku} {and flat-band superconductivity \cite{koba}} become possible at high temperature. Exotic heavy excitons may exist \cite{Chamon}. In an external magnetic field, Landau levels of charged fermions are proportional to the square root of odd integers 
 while there are huge number of flat bands corresponding to Landau indices \cite{aoki,Goldman}.  A large incident angle Klein tunneling was predicted \cite{Xing}. {Though the chirality of Dirac-Weyl fermions with integer spin can be defined \cite{chi1,chi2},} it was asserted that there is no chiral anomaly because the Dirac-like operator is not  an elliptical operator \cite{Lan}. The  dc conductivity will diverge due to the flat band  \cite{Vigh}.

In this article, we would like to point out  that the ferotons with a zero flat band are of the fermionic gauge symmetry in analog to the electromagnetic gauge symmetry in Maxwell theory and have the same longitudinal component {as photon}.  The exact zero flat bands may be redundant gauge degrees of freedom, similar to the longitudinal photon in Maxwell theory. 
{Because photon is neutral, it does not couple to electromagnetic field. The breakdown of electromagnetic gauge symmetry arises from the Higgs mechanism in superconductivity, i.e., the condensation of cooper pairs of electrons. Losing gauge symmetry generates a mass to the longitudinal photon which becomes physical and causes Meissner effect in superconductors.} Different from the photon, the feroton is a quasiparticle in an electron system and then is charged. It may couple to the electromagnetic field. This coupling leads to the breakdown of the fermionic gauge symmetry and these fermionic gauge degrees of freedom may turn to be physical. Picking back  these degrees of freedom will cause interesting emergent phenomena.

 In an external magnetic field, we {analytically} calculate the Landau level spectra of the feroton. Besides a chiral gapless mode in the zeroth Landau level which arises from the chiral anomaly \cite{NN}, there are many {other non-topological gapless modes. These gapless modes are characterized by Landau level indices ($n>1$) and their chiralities in the long wave length limit are all opposite to the chiral anomaly mode.} Without topological protection, the chiralities of these gapless modes may change signs at a finite wave vector. The energies of these modes tend to zero as the wave vector goes to infinity.  The band widths of these modes are very narrow. {If we turn on an extremely small electric field $E$ parallel to the magnetic field, then most of the gapless modes will contribute to the conductance of the feroton semimetal. The magnetoresistance of the feroton semimetal approaches $-1$ as that in a metal.} As $E$ increases, the modes with narrower band widths {which are fully below the chemical potential} will not  participate in the transport. {As $E$ increases further, fewer and fewer gapless modes contribute while the chiral anomaly induced zero Landau level band becomes dominating.} This leads to a giant positive magnetoresistance first and then to a giant negative one as $E$ increases.  Finally, all gapless modes {are below the chemical potential except for the chiral anomaly induced zero Landau level band}, which gives a large negative magnetoresistance similar to that in the WSMs.

At a finite Fermi energy, the transverse ferotons of the feroton semimetal in an external field exhibit a quantum oscillation of the DOS as that in WSMs \cite{qo}. If the flat band is not exact flat, say perturbed by a quadratic dispersion with a mass $m^*$, the gapless modes may be gapped by  additional Landau levels of the quadratic term.  An extra low energy quantum oscillation corresponding to the longitudinal feroton may appear.  Since in the latter case, the Fermi level is very close to the triply degenerate nodal point, the length of the measured Fermi arc may be closer to the linear distance between two triply degenerate nodal points.  This may be confirmed by  numerical calculations for a lattice model on a three dimensional generalized Lieb lattice. To explore more exotic effects from the gapless modes, we also study the ferotons on a cubic lattice where the perturbed quadratic term near a triply degenerate nodal point is two dimensional. In this case, there is no Landau gap between the  $n>1$ low energy Landau bands.  An extra quantum oscillation of the DOS may still be observed.  

To show the feroton semimetal possibly emerges in real materials, we {check up} the existed {\it ab initio} calculations for the band structures of  several known and proposed compound materials. The evidences of the feroton semimetal are revealed in Pd$_3$Bi$_2$S$_2$ and Ag$_3$Se$_2$Au \cite{ber}, LPtBi(L=La,Yp,Gd) \cite{tdnp1}  and ZrTe \cite{tdnp4}.

This paper is organized as follows: In Sec. II, we discuss the similarity between the flat band of the ferotons and the longitudinal photon. We show the fermionic gauge symmetry {of the flat band of feroton}. In Sec. III, we propose a generalized feroton theory and the corresponding lattice model, a three-dimensional generalized Lieb lattice model. {There are eight feroton points at the high symmetry points in the Brillouin zone, each of which has a definite chirality corresponding to its monopole charge.} In Sec. IV, {we couple the feroton to an external magnetic field and this breaks the fermionic gauge symmetry. We calculate the Landau levels analytically and we find that the original flat band becomes gapless narrow bands labeled by Landau level indices.} In Sec. V, we study the anomalous magnetic transport property caused by the {gapless} longitudinal ferotons. In Sec. VI,  we show two quantum oscillations of the DOS at different energy scales. In  a special cubic lattice, the extra quantum oscillation may happen even without low energy Landau level gaps. In Sec. VII, we discuss {some potential material realizations} in which the ferotons may emerge. The final section is {devoted to} our conclusions.

\section{Fermionic gauge symmetry}

\subsection{ "Primordial flat bands" in gauge fields}

The history of research on the flat bands  in condensed matter systems  casts back to 1970s in amorphous semiconductors \cite{wea,weath,thwea}.  The {emergence of the} flat band means there are extremely strong correlations in many body  systems \cite{esc1,esc2}. {Besides the condensed matter cases,} 
there is a much earlier physical theory {with a flat band}, the Maxwell theory of the electromagnetic field, in which the longitudinal photon is dispersionless and so is of a "primordial flat band" for a photon gas.  The Maxwell equations read
$$
(\partial^2g_{\mu\nu}-\partial_\mu\partial_\nu)A^\nu=0,
$$
where $\mu,\nu=0,x,y,z$ and $g_{\mu\nu}={\rm diag }(1,-1,-1,-1)$.
Taking $A_0=0$,  the equations of motion of $A_i$ with the constraint ${\cal E}\nabla\cdot{\bf A}=0$ are given by
 \begin{equation}
		( -\nabla^2\delta_{ij}+\partial_i\partial_j)
	A_j={\cal E}^2A_i, \label{MW}
\end{equation}
where ${\cal E}$ is the energy of the gauge field.
We here {not} yet considered the residual gauge symmetry.
It is easy to check that $A^{(0)}_\mu({\bf r})=(0,\partial_i \Lambda({\bf r}))$ with  arbitrary  time-independent scalar function $\Lambda({\bf r})$ is a zero energy solution. In the momentum space, $A^{(0)}_\mu({\bf q})=(0,q_i\Lambda({\bf q}))$ is an exact flat band.
This zero energy flat band is identified as the dispersionless longitudinal photon which is unphysical redundant degrees of freedom {and generates the gauge transformations}.  In fact, $A_i({\bf r},t)\to A_i({\bf r},t) +(0,\partial_i \Lambda({\bf r}))$ is the residual time-independent gauge transformation in $A_0=0$ gauge.

A massless higher-spin matter field can also transform as a gauge field \cite{RS}.  For example, for spin-$\frac{3}2$ relativistic  spinor-vector $\Psi_{\nu\alpha}$ \cite{note},  Rarita and Schwinger in 1941 recognized that the Lagrangian
$$
L_{\rm RS}=\bar{\Psi}_\mu((g^{\mu\nu}\gamma \cdot \partial)-(\gamma^\mu\partial^\nu+\partial^\mu\gamma^\nu)+\gamma^\mu(\gamma\cdot\partial)\gamma^\nu)\Psi_\nu
$$
is  invariant under the gauge transformation $\Psi'_{\mu\alpha}=\Psi_{\mu\alpha}+\partial_\mu\eta_\alpha$ \cite{RS}. Here, $\gamma_\mu$ are gamma matrices. Therefore, the gauge invariance under a static gauge transformation $\delta \Psi^{(0)}_{\mu\alpha}=(0,\partial_i \eta_\alpha({\bf r}))$  implies  that $\delta \Psi^{(0)}_{\mu\alpha}({\bf q})$ correspond to  four zero energy flat bands.  These flat bands are redundant degrees of freedom.  The genuine physical degrees of freedom are the Rarita-Schwinger-Weyl fermions with the highest and lowest helicity \cite{Lure}. In fact, any higher spin relativistic massless particles are of such a gauge symmetry \cite{RS}. The physical degrees of freedom are the highest and lowest helicity gapless modes \cite{Lure}.  

\subsection{Flat band with zero energy in spin-1 gauge fermions}

 Keeping the redundant flat band {interpretation} of the longitudinal  photon and gauge fermions in mind, we {propose another possible gauged flat band condensed matter system, i.e., the pseudospin one fermion with linear dispersion and a flat band. Since the constitution particles are non-relativistic fermions and bosons in condensed matter systems, the quasiparticles are not subject to Lorentz symmetry and the spin-statistics relation can be violated \cite{ssc,ssc1}.} The pseudospin one fermions are frequently investigated \cite{Lieb,Berc}. In three dimensions, the  Hamiltonian is given by \cite{ber}
 \begin{eqnarray}
 H_{\rm s1}={\bf q}\cdot {\bf S}, \label{s1} 
 \end{eqnarray}
where $S^i~(i=1,2,3)$ are $3\times3$ matrices of SO(3) generators. For convenience,  we take { $S^i_{jk}=-i\epsilon_{ijk}$ with $jk$ labeling the matrix elements of $S^i$. }
The Sch\"odinger equations are  
\begin{eqnarray}(H_{sl})_{ij}\psi_j=E\psi_i \label{sls}
\end{eqnarray}
and the eigen solutions are well-known: There are two bands $\boldsymbol{\psi}^{(\pm)}$ with the eigenenergies $\pm q$ and a flat band $\boldsymbol{\psi}^{(0)}$ where 
\begin{eqnarray}
\boldsymbol\psi^{(\pm)}=\frac{1}{\sqrt {2(q_1^2+q_2^2)} q}\left(
\begin{array}{ccc}
 -(q_3q_1\mp iqq_2) \\
 -(q_3q_2\pm iqq_1)  \\
 q_1^2+q_2^2
\end{array}
\right),\nonumber
\end{eqnarray}
and $\boldsymbol{\psi}^{(0)}={\bf q}/q$. The wave functions $\boldsymbol\psi$ of the ferotons are also the 
solutions of Eq. (\ref{MW}), {which correspond to the transverse photon with energy $q$ and the longitudinal photon respectively.}  
 In analog  to the Eq. (\ref{MW}),  the Schr\"odinger equations are gauge invariant under $\delta \boldsymbol{\psi}={\bf q}\Lambda({\bf q})$ which is the flat band zero solution. If the gauge symmetry was not broken, the flat zero mode would be redundant. Actually,  if we define the static transverse  projection $P_T=\sum_{\sigma=\pm}\psi^{(\sigma)}_i\psi^{(\sigma)}_j=
 \delta_{ij}-\frac{q_iq_j}{q^2}$, the flat band is the longitudinal feroton while $\boldsymbol{\psi}^{(\pm)}$ are the transverse ferotons, similar to their bosonic counterparts, the photons. The gauge degrees of freedom $\delta \boldsymbol{\psi}={\bf q}\Lambda({\bf q})$ belong to the space of states $P_L\psi=\psi_L$ with $P_L=\psi^{(0)}_i\psi^{(0)}_j=\frac{q_iq_j}{q^2}$. Namely, the ferotons are a three component gauge field. To distinguish with the electromagnetic gauge symmetry, we call this gauge symmetry the fermionic gauge symmetry.
 
 Similar to the Weyl fermion, the transverse degrees of freedom gives a nontrivial Berry curvature
 \begin{eqnarray}
 \vec{\cal B}=\nabla_q\times(\boldsymbol{\psi}^{(\pm)*}\cdot\nabla_q \boldsymbol{\psi}^{(\pm)})=2\frac{\bf q}{q^3}.
 \end{eqnarray}
 This implies the feroton node may be thought of as a monopole with charge 2 in the momentum space, 
{ the same as a double Weyl node \cite{doubleweyl}. But the dispersion of a double Weyl node cannot be linear in all three directions which distinguishes the double Weyl node with the feroton node. Thus, the feroton cannot be identified to any kind of Weyl fermion. Besides the nontrivial monopole charge, the three dimensional crossing of three levels for the feroton is stabilized by crystal symmetry, especially, there are three dimensional representations { for triply-degenerate points} at high symmetry points with space group symmetry 199, 214 and 220 \cite{ber}. }

\subsection{The physical degrees of freedom }

The longitudinal photon in vacuum or a normal matter cannot be detected. However, if the gauge symmetry is broken, the gauge degree of freedom becomes physical. The Meissner effect in superconductors is the result of gauge symmetry breaking which gives the longitudinal photon a finite mass.

The fermionic gauge invariance of the Schr\"odinger equation (\ref{sls}) indeed implies that the flat band with zero energy or flat zero mode at the Fermi level is a redundant gauge degree of freedom. {By breaking the fermionic gauge symmetry, these redundent degrees of freedom can be recovered.} Some examples are as follows: 

(i) Generally speaking, a flat band is not a redundant  gauge degree of freedom because the chemical potential term added to the Hamiltonian (\ref{s1}) breaks the gauge symmetry of Eq. (\ref{sls}). A finite energy flat band has interesting physical effects, e.g., the peculiar Klein tunneling for the spin-1 fermion \cite{Xing}. 
(ii) Because the corresponding "Dirac"  operator corresponding to the Hamiltonian (\ref{s1}) is the curl operator which is not elliptic,  the chiral anomaly does not exist in the spin-1 system with the Hamiltonian (\ref{s1}). If the flat band is lifted to a finite energy, the zero modes in the physical space are finite. The transverse ferotons carry helicity $\pm1$ which give the Chern numbers $\pm2$ {for any surface containing the feroton point with $q_3\ne 0$.} The chiral anomaly is given by this Chern number and there is a quantum anomalous Hall effect with $C=2$ in two-dimensions. The Fermi arcs on the surface of the system exist, similar to the ones in the Weyl semimetal \cite{wan}. (iii) {Because the feroton quasiparticles carry charges, they can couple to electromagnetic fields. Coupling to an external electromagnetic field will break the fermionic gauge symmetry of the feroton. Under a magnetic field, the gauged flat band degrees of freedom become physical, dispersive and gapless. These gapless Landau levels will produce novel magnetic transport properties of a feroton semimetal which we will discuss in details later.}

\section{Generalized feroton  model }

\subsection{Generalized ferotons }

{In this subsection, we present a generalized feroton model which hides a local fermionic gauge symmetry. The Hamiltonian we consider reads,} 
\begin{eqnarray}
H_{\rm GFB} &=&v_1{\bf  q}\cdot{\bf S}
+v_4q^2 +(v'_4-v_4)(q^2_i\delta_{ij})_{3\times 3}\nonumber\\
&-&v_2 (q^2_i\delta_{ij}-q_iq_j)_{3\times3}. \label{GFB}
\end{eqnarray}
The coefficient choices are for latter convenience.
We consider two special cases with flat bands. First, we take $v_2=-v_4,v_4'=0$, the Hamiltonian (\ref{GFB}) is reduced to
\begin{eqnarray}
H^T_{\rm GFB}=v_1{\bf q}\cdot{\bf S}+v_4(q^2\delta_{ij}-q_iq_j)_{3\times3},
\end{eqnarray}
where {the first term is Chern-Simons like and} the second term is just the Maxwell Hamiltonian in (\ref{MW}) but acting on the fermions. Then, the transverse modes $\boldsymbol\psi_\pm=\frac{1}{\sqrt2 q}((q_xq_z\mp iqq_y)/\sqrt{q_x^2+q_y^2},(q_yq_z\pm iqq_x)/\sqrt{q_x^2+q_y^2},-\sqrt{q_x^2+q_y^2})$ have the dispersions
\begin{eqnarray}
E_\pm=\pm v_1 q-v_4q^2.
\end{eqnarray}
 The longitudinal mode $\boldsymbol\psi_0= {\bf q }\Lambda({\bf q})$ is a flat band, which is redundant and must be gauge away. This is the transverse feroton model.

  The second case is to take $v_1=v_4=0$ and $v'_4=-v_2$ and then the Hamiltonian (\ref{GFB}) is reduced to the longitudinal part
 \begin{eqnarray}
 H^L_{GFB}=-v_2(q_iq_j)_{3\times3}.\label{LGH}
 \end{eqnarray}
 The longitudinal mode has a dispersion $E_0=-v_2q^2$ while two transverse modes $\boldsymbol{\psi}_\pm\Lambda_\pm({\bf q})$ are flat bands which must be gauged away.  This is the longitudinal feroton model.

\subsection{ Lieb lattice in 3 dimensions }

\begin{figure}[ptb]
\centering
\includegraphics[width=7.2cm]{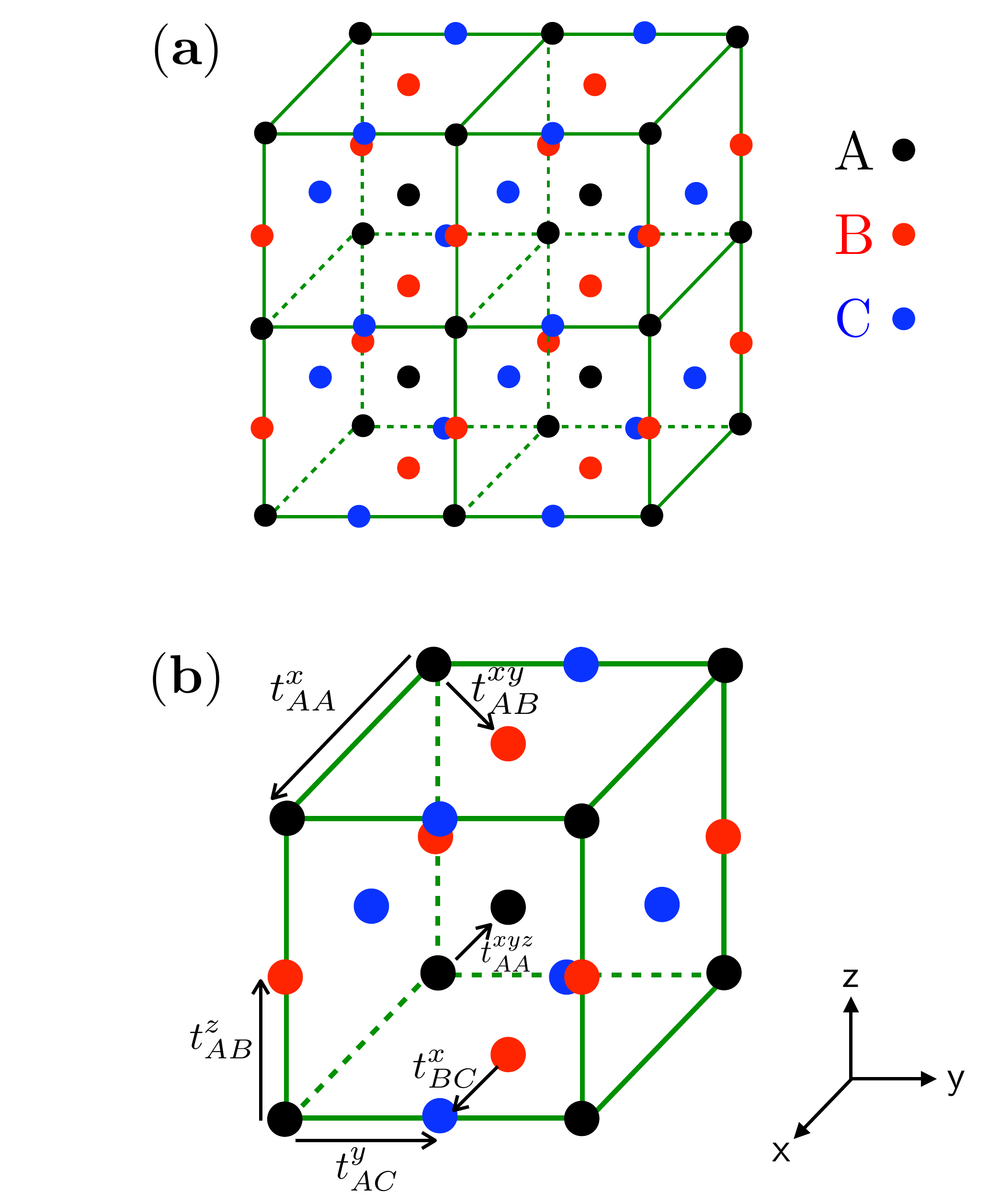}
\newline\caption{The three-dimensional generalized Lieb lattice. (a) Lattice structure with three body-centered cubic sublattices coupled together. The distance between the nearest neighbors is set to 1. (b) Typical hoppings. The nearest and the second nearest neighbor hoppings are taken as $t^z_{AB}=t_{AC}^y=t_{BC}^x=-t^{-z}_{AB}=-t_{AC}^{-y}=-t_{BC}^{-x}=t_1$ and $t^{xy}_{AB}=t^{-x,-y}_{AB}
=t_{AC}^{xz}=t_{AC}^{-x,-z}=t_{BC}^{yz}=t_{BC}^{-y,-z}=-t^{-x,y}_{AB}=-t^{x,-y}_{AB}
=-t_{AC}^{-x,z}=-t_{AC}^{x,-z}=-t_{BC}^{-y,z}=-t_{BC}^{y,-z}=t_2$.
The third nearest neighbor hoppings, for example $t_{AA}^{xyz}$, are all taken as $t_3$. The fourth nearest neighbor hoppings are defined as $t_{AA}^{y}=t^{z}_{AA}=t_{BB}^{x}=t^{z}_{BB}=t_{CC}^{x}=t^{y}_{CC}=t_4$ and $t_{AA}^{x}=t_{BB}^{y}=t^{z}_{CC}=t'_4$.}
\label{fig1}
\end{figure}

\begin{figure}[ptb]
\centering
\hspace{0.4cm}
\includegraphics[width=5.6cm]{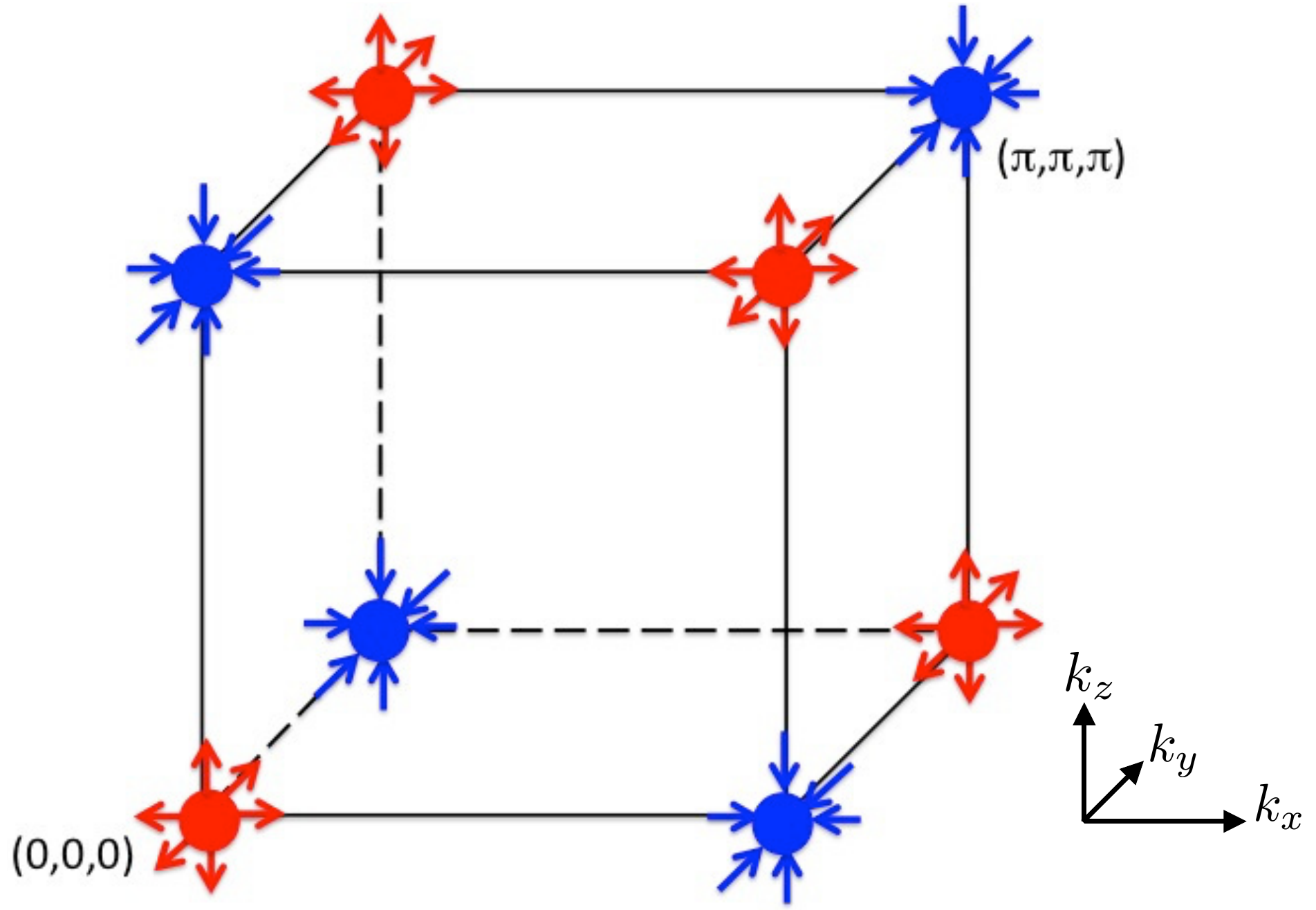}
\newline\caption{{The nodal points of the Lieb lattice model. The red nodes with outgoing arrow carry a monopole charge 2 while the blue nodes with ingoing arrow carry a charge -2. } }
\label{fig22}
\end{figure}

We construct a  Lieb lattice model on cubic lattice whose long wavelength limit at $\Gamma$ point {reduces to} Eq. (\ref{s1}).  
The tight-binding model of spinless fermions includes the first, second, third and fourth nearest neighbor hoppings. (See Fig. \ref{fig1}.) We take $t_3=0$ for simplicity.  After properly adjusting the chemical potential and shifting the zero energy to the Fermi level, the Hamiltonian is given by
\begin{eqnarray}
H_{\rm GL} &=&-2t_1\boldsymbol{\kappa}\cdot{\bf S}
-4t_4\kappa^2 -4(t'_4-t_4)(\kappa^2_i\delta_{ij})_{3\times 3}\nonumber\\
&+&4t_2 (\kappa_i^2\delta_{ij}-\kappa_i\kappa_j)_{3\times3}, \label{GL}
\end{eqnarray}
where $\boldsymbol{\kappa}=(\sin q_x, \sin q_y, \sin q_z)$. 
Obviously, at the $\Gamma$ point, $H_{\rm GL}$ recovers (\ref{GFB}). There are eight nodal points located at $(0,0,0)$, $(\pi,0,0)$, $(0,\pi,0)$, $(0,0,\pi)$, $(\pi,\pi,0)$, $(\pi,0,\pi)$, $(0,\pi,\pi)$ and $(\pi,\pi,\pi)$ (see Fig.~\ref{fig22}). For $t_2=t_4$ and $t_4'=0$, the monopole charges of the corresponding nodes are $\pm 2$, as illustrated in Fig. \ref{fig22}. {The sign of the linear part in the effective Hamiltonian at each nodal point is determined by its monopole charge.}

 The model (\ref{GL}) can also be reduced to two dimensions. 
 There were plenty of works in this Lieb lattice model as mentioned in Introduction.  The Hamiltonians in those works are similar to the two-dimensional reduction of (\ref{s1})  up to a unitary transformation.  A similar two-dimensional model can also be realized on ${\cal T}_3$  lattice \cite{Berc}.

\subsection{Nearly flat band}

The Hamiltonian  (\ref{s1}) gives an exact flat band. In a real system, {the exact flat band usually does not exist.  There are several situations of breaking the flat band gauge symmetry}: (i) If the nearly flat band can be decomposed into an exact flat part plus a perturbation, the exact flat degrees of freedom are still gauged ones, i.e., the unperturbed Hamiltonian is the fermionic gauge invariant  which is perturbatively broken.
(ii) There is no such a decomposition for the nearly flat band. For example, if the nearly flat band in two dimensions are topologically nontrivial, it cannot  continuously tend to exact flat \cite{CLi}. However, in three dimensions, there are topologically nontrivial exact flat bands. The transverse modes for the Hamiltonian (\ref{LGH}) are examples. They are exact flat but their monopole charges are $\pm2$. Reducing to two dimensions, $\boldsymbol\psi_\pm$ are topologically trivial, in consistent with the theorem proved in \cite{CLi}. (iii) The fermionic gauge  symmetry is nonperturbatively broken. For instance, {applying an external magnetic field will break the fermionic gauge symmetry which we will study right below.}

\section{Landau levels of ferotons}

The large negative magnetoresistance is a ubiquitous property in topological insulators, Dirac and Weyl semimetals, stemming from the chiral anomaly \cite{NN}.  These phenomena were observed experimentally \cite{Kim,LiSY,Huang}. 
{In this three component fermion model, the existence of electromagnetic field not only provides such an anomalous electric transport phenomenon through the chiral anomaly in the lowest Landau level \cite{ber}, but also breaks the fermionic gauge symmetry and induces gapless Landau levels with narrow bands. Therefore the electric transport phenomena for the pseudospin one feroton are expected to be much richer than those in the spin-1/2 systems.}

\begin{figure}[ptb]
  	\centering
  	\includegraphics[width=8.6cm]{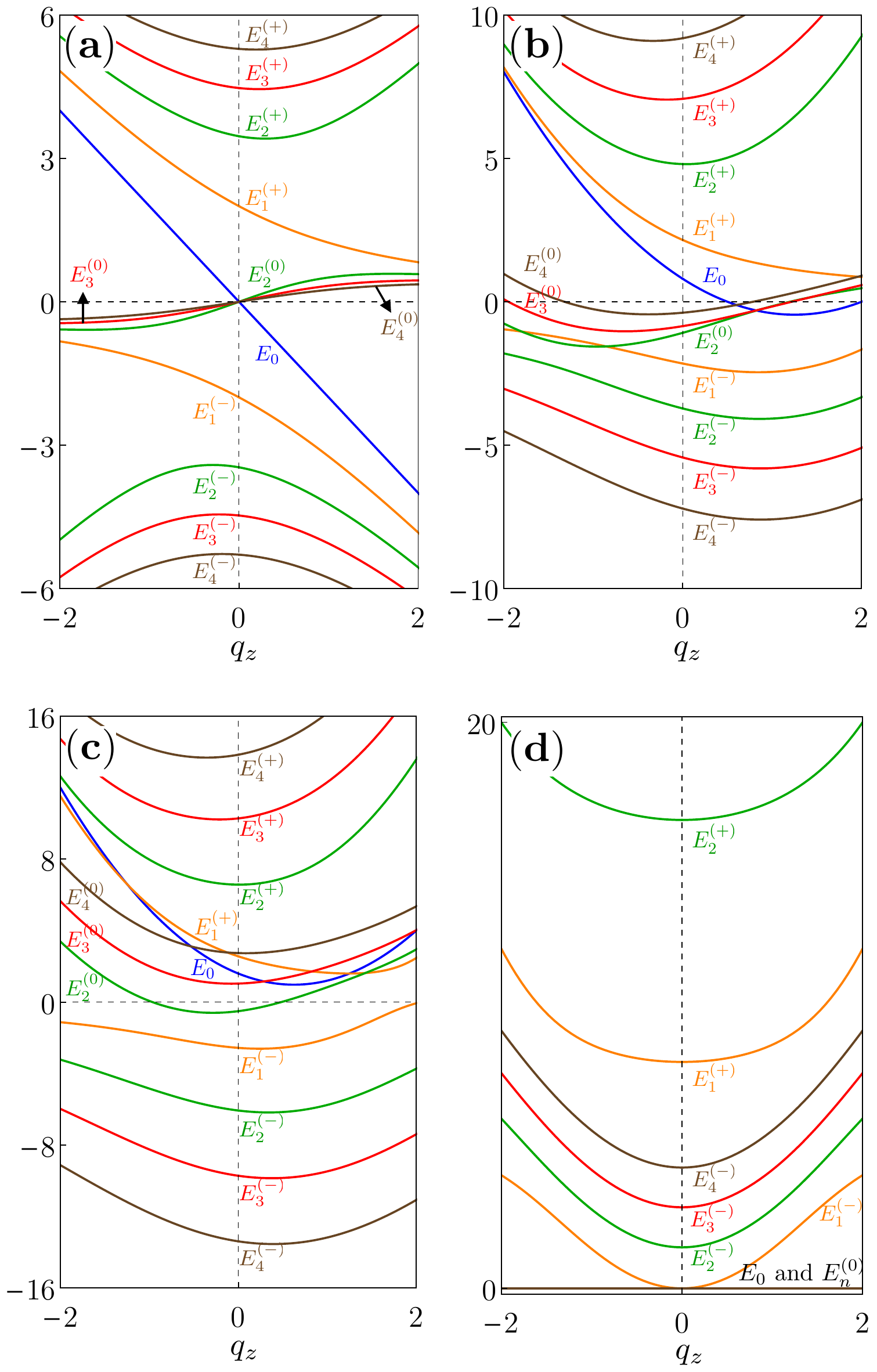}
  	\newline\caption{{The blue, orange, green, red and brown curves represent the Landau levels with index $n=0,1,2,3,4$, respectively.} The parameters are chosen as (a) $v_1=-2, v_2=v_4=0$; (b) $v_1=-2, -v_2=v_4=0.8$; (c) $v_1=-2, -v_2=v_4=1.6$; (d) $v_1=v_4=0, v_2=2$.}
  	\label{fig2}
  \end{figure}

 We consider the Hamiltonian (\ref{GFB}) for the charged fermions in an external magnetic field  applied in the $z$ axis. For simplicity, we take  $v_4'-v_4=v_2$. The Hamiltonian then reads
 \begin{eqnarray}
 H_{M}=v_1{\bf S}\cdot{\bf D}+v_4D^2+v_2(D_iD_j)_{3\times3}, \label{EM}
 \end{eqnarray}
where $D_a=-i\hbar\partial_a+\frac{e}c A_a$ ($a=1,2$) with the magnetic field ${\bf B}=\nabla\times {\bf A}$ in the $z$ direction while $D_3=q_z$ since there is no $z$-dependence of ${\bf A}$ and $A_3=0$. We take $\hbar=e=c=1$ and the magnetic length $l_B=\sqrt{1/B}=1$. Due to $[D_1,D_2]=iB=i$, the Hamiltonian (\ref{EM}) is not fermionic gauge invariant under $\delta\boldsymbol{\psi}={\bf D}\Lambda$ {which no longer generates the zero energy solutions} even if $v_2=-v_4$ (Do not confuse with the electromagnetic gauge transformation).
Thus, the exact flat band disappears and the redundant fermionic gauge degrees of freedom are converted into physical ones due to the breaking of the {fermionic} gauge symmetry.
The Hamiltonian (\ref{EM}) can be analytically diagonalized.  Through a basis rotation $\Phi_{1,3}=\frac1{\sqrt2}(i\psi_2\pm \psi_1), \Phi_2=\psi_3$,  the Hamiltonian $H_M$ transforms to
\begin{eqnarray}
\left(
\begin{array}{ccc}
	I_1
	& (-v_1+v_2q_z)a^\dagger
	&-v_2a^{\dagger 2} \\
	(-v_1+v_2q_z)a
	& I_2
	& (-v_1-v_2q_z)a^\dagger \\
	-v_2a^2
	& (-v_1-v_2q_z)a
	& I_3
\end{array}
\right),\label{TH}
\end{eqnarray}
where $a$ and $a^\dag$ obeying $[a,a^\dag]=1$ are the Landau level lowing and raising operators, $I_1=(2v_4+4v_2)a^\dagger a+v_4+v_1q_z+v_4q_z^2$, $I_2=2v_4a^\dagger a+v_4+(v_4+v_2)q_z^2$ and $I_3=(2v_4+v_2)a^\dagger a+(v_4+v_2)-v_1q_z+v_4q_z^2$. Denoting the $n$-th Landau level wave functions by $\varphi_n$ 
{with the degeneracy $\Phi/\Phi_0$ where $\Phi$ is the total magnetic flux through the system and $\Phi_0$ is the flux quanta.} The general solutions of the problem are $\boldsymbol{\Psi}_n=(C_1\varphi_n,C_2\varphi_{n-1},C_3\varphi_{n-2})$ where $\varphi_{-1,-2}$ are taken to be 0. For $n=0$, there is only one Landau band with $E_0=v_4+v_1q_z+v_4q_z^2$, which is a chiral mode in a chain along the $z$-direction when $v_1^2-4v_4^2>0$ (See Figs. \ref{fig2}(a) and (b)) and {originated} from the chiral anomaly as that for the WSM. The chiral anomaly is destroyed by the larger quadratic term if $v_1^2-4v_4^2<0$. See Fig. \ref{fig2}(c) where the {original chiral mode disappears and minimizes above the zero energy.}  For $n=1$, there are two Landau bands as shown in Fig. \ref{fig2} and they also relate to the chiral anomaly \cite{ss,beenkker}.

For a given $n>1$, three Landau bands are obtained by solving the eigen equation of (\ref{TH}). For $v_2=v_4=0$, {the zeroth and the first Landau level has the dispersion $E_0=v_1q_z$, and $E_1^{(\pm)}=v_1(q_z\pm\sqrt{q_z^2+4B})$. The robustness of these two Landau levels is guaranteed by the chiral anomaly. For a general n($\geqslant 2$)th Landau level, there are three eigenvalues with 
\begin{eqnarray}
E_n^{(+)}&=&\omega_1(q_z/2+\sqrt{\Delta})^{\frac{1}{3}}+\omega_2(q_z/2-\sqrt{\Delta})^{\frac{1}{3}},\\
E_n^{(-)}&=&\omega_2(q_z/2+\sqrt{\Delta})^{\frac{1}{3}}+\omega_1(q_z/2-\sqrt{\Delta})^{\frac{1}{3}},\\
E_n^{(0)}&=&v_1((q_z/2+\sqrt{\Delta})^{\frac{1}{3}}+(q_z/2-\sqrt{\Delta})^{\frac{1}{3}}),
\end{eqnarray}	
where $\omega_{1(2)}=t_1(-1\pm\sqrt{3})$ and $\Delta=q_z^2/4-(2n+1+q_z^2)^3/27$. From the above energy spectrum, we conclude that }for a given $n>1$, two bands are gapped with energies which are $\pm v_1\sqrt{2n+1}$ at $q_z=0$ and there is a gapless mode whose chirality at $q_z=0$ is opposite to the chiral mode of the zeroth Landau level  \cite{note1}. The slope of the gapless mode at $q_z=0$ is $\frac{-v_1}{2n+1}$. The chirality of these gapless modes is not topologically protected but stems from the breaking of the fermionic gauge symmetry. Fig. \ref{fig2}(a) shows the spectra {from the zeroth Landau level to the fourth one for this case. The Landau bands with $n>4$ are not illustrated but they are similar to that of $n=2,3,4$.}  The band widths of these gapless modes 
{are proportional to $4v_1/\sqrt{2n+1}$ and become }very narrow as $n\gg 1$. The energies of these {emergent gapless} modes eventually tend to zero when $q_z\to \infty$.  Fig. \ref{fig2}(b) depicts the results with the quadratic perturbation, which are similar to that with $v_2=v_4=0$. As the quadratic terms dominate, the dispersions are expected to close to parabolics as shown in Fig. \ref{fig2}(c).

For the longitudinal gauge fermion model ($v_1=0$, $v_2=v_4'$ and $v_4=0$), there is only an $E=0$ band for $n=0$. For $n=1$, two bands are given by 
$$E^2-v_2(2+q_z^2)E+v_2q_z^2=0.$$
 For $n>1$, there is always an $E=0$ band and the other gapped two are given by 
 $$E_n^{(\pm)}=(2n+q_z^2)v_2/2\pm v_2\sqrt{4(n^2+n-1)+4nq_z+5q_z^2}/2.$$ These dispersions are plotted in Fig. \ref{fig2}(d).  
 {Notice that the flat band here cannot be gauged away because the fermionic gauge symmetry is broken by the external electromagnetic field. To be more precise, after a gauge transformation, namely, by multiplying an arbitrary $U(1)$ phase to the flat band wave function, the energy of the transformed wave function is longer zero.}
We will investigate this longitudinal gauge fermion elsewhere for details.

\begin{figure}[ptb]
\centering
\includegraphics[width=7.2cm]{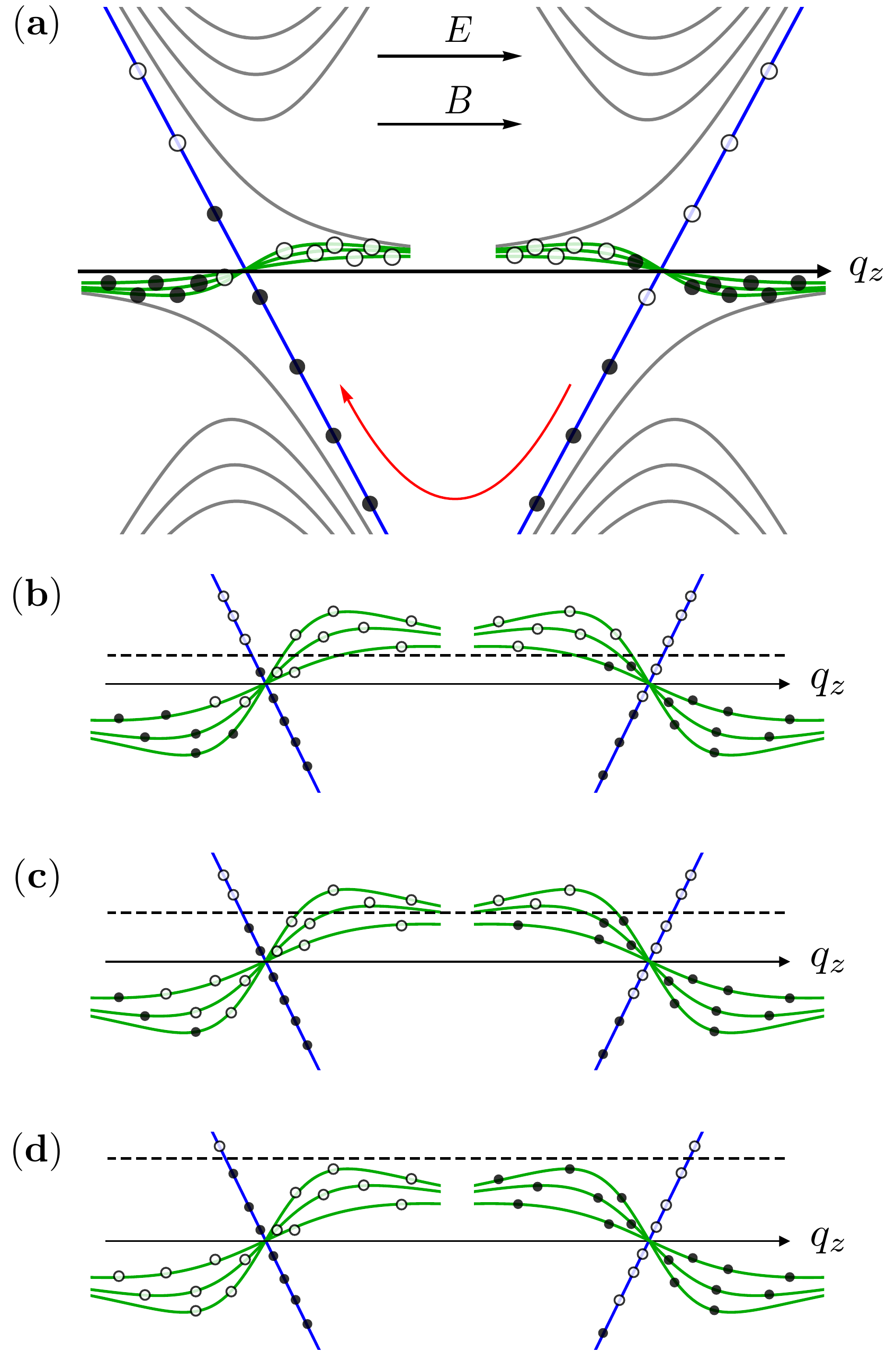}
\newline\caption{The quantum charge pumping between Weyl nodes in parallel electric and magnetic fields. 
		(a) We adapt Fig.~4 in Ref. \onlinecite{qi} to sketch the quantum charge pumping between two paired ferton points in our case. 
		The blue bands are the lowest Landau level chiral modes, the green bands 
		are the chiral modes of higher Landau levels (only a few of them are depicted) and the gray bands
		are irrelevant (gapped) modes. The black (white) dots indicate occupied (unoccupied) states.
 		(b)-(d) We focus on the relevant (gapless) modes. The dashed lines are the chemical potential of the left-moving fermions due to the variation of the $E$ field. (b) In a very weak $E$ field, a large amount of chiral modes contribute to the electric current. (c) A few modes, say between 10 to 20, contribute. (d) Only the lowest Landau level chiral modes do.
}

\label{fig3}
\end{figure}

 \begin{figure}[tb]
\centering
\includegraphics[width=5.8cm]{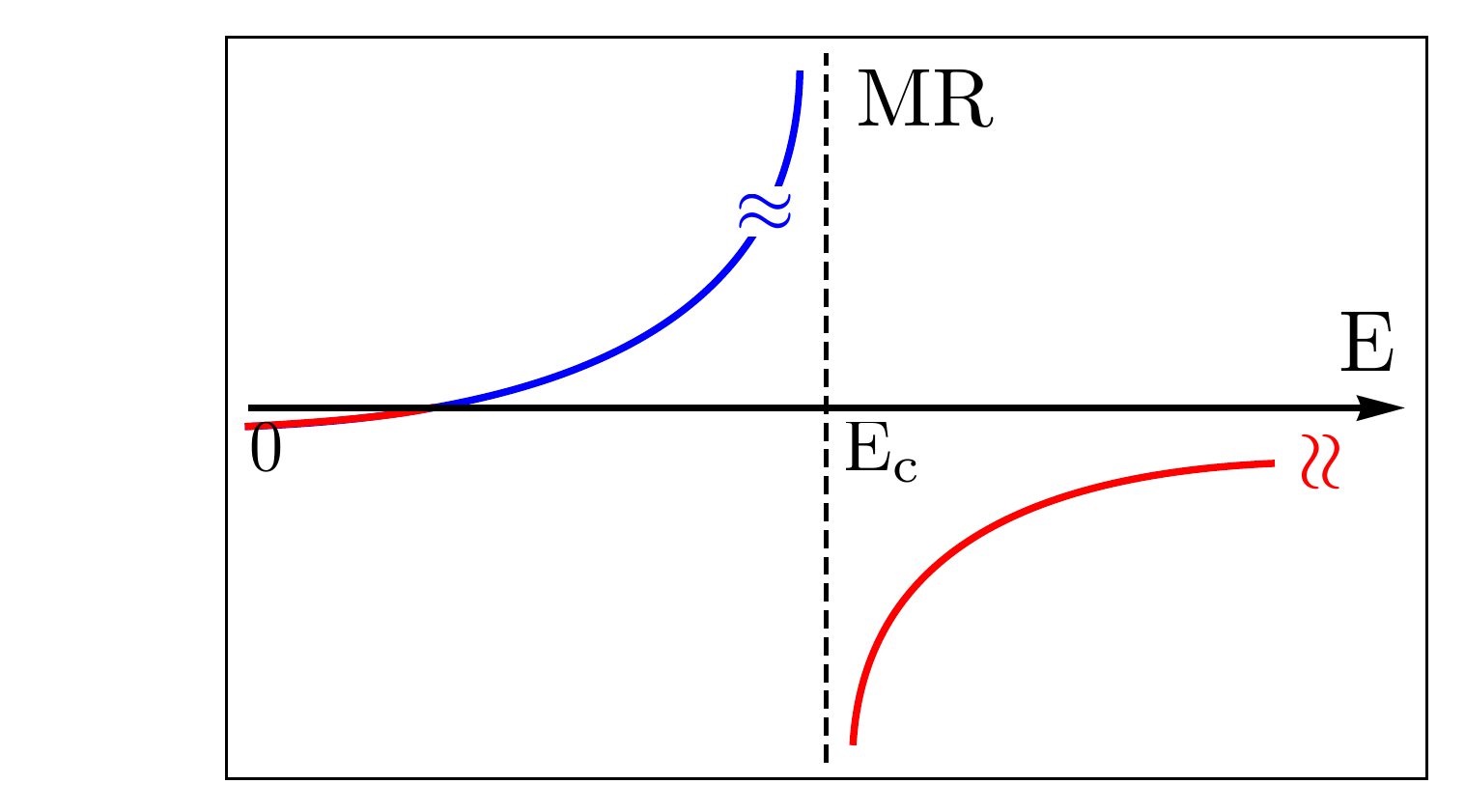}
\newline\caption{The schematic of the magnetoresistance. Positive (negative) magnetoresistance is indicated by blue (red). There is a discontinuous change when crossing a critical E field $E_c$, see main text.}
\label{fig4}
\end{figure}

\section{Magnetic transport property}

 The giant negative magnetoresistance is a significant phenomenon in topological semimetals \cite{qi,yan}.   Due to the existence of the new emergent gapless modes, combining with the topological chiral mode, the transport phenomena are much richer. We focus on the model with the spectrum of Fig. \ref{fig2}(a) and consider the transport around the zero energy. For a given $n>1$, 
{the maximum and minimum of the band energies of the gapless modes scale as $\pm|\frac{v_1}{\sqrt{n+1/2}}|$} and tend to zero as $1/q_z$ when $|q_z|\to \infty$. We notice that all these gapless modes are orthogonal to each other. Especially, the backward scattering between the chiral anomaly mode and the other modes from the breakdown of fermionic gauge symmetry  is forbidden.  
 
 {After applying an electric field along the direction of the magnetic field, the quantum charge pumping  
 between two feroton nodal points with opposite monopole charges, for instance,} $(0,0,0)$ and  $(0,0,\pi)$, is similar to  that between two Weyl points \cite{qi}. The effect of the electric field can be understood in low energy "quantum limit"  \cite{qi}, {see Fig.~\ref{fig3}}. For ${\bf E}\parallel {\bf B}$, most of the states move along the electric field according to $\hbar\dot{\bf q}_n=-e_n{\bf E}$ .  At a given nodal point, since the zeroth Landau level is chiral and the low energy $n>1$ modes in the long wave length limit are of the opposite chirality,  $e_{n>1}=-e_0=-e$ at $\Gamma$ while $e_{n>1}=-e_0=e$ at $(0,0,\pi)$. For $n=0$, the charge $-e$ ferotons are pumped from $(0,0,\pi)$ to $(0,0,0)$ while the charge $-e$ ferotons are pumped from from $(0,0,0)$ to $(0,0,\pi)$ for $n>1$ bands.  If we apply a very weak electric field $E$ along the $z$-direction such that the chemical potential difference $\Delta\mu$ between the left- and right-moving ferotons is of the order $|v_1|/\sqrt {n_{max}}$ with a very large $n_{max}$.  The $n\leq n_{max}$ gapless modes as well as the chiral anomaly mode participate in the charge transport ({See Fig. \ref{fig3}(b)}). The reason that the zero modes with $n>n_{max}$ do not contribute to the current is as follows: {As seen in Fig. \ref{fig3}(c)}, when a chiral mode with its maximal energy is smaller than $\Delta\mu/2$, all charges in the left Weyl point are pumped to the right Weyl point. Since the filled gapless band includes both states with positive and negative velocities, the contributions of all states to the current cancel. 
{For a strong magnetic field B, the chemical potential $\mu<<\hbar \omega_c$ where $\omega_c$ is the cyclotron frequency. Under a parallel electric field E, the current in the quantum limit is given by (the details are presented in Appendix \ref{app1}),
	\begin{eqnarray}
	j_z&=&j_z^0+\sum^{n_{max}}_{n=2}j_z^n=\left(1-\sum_{n=2}^{n_{max}}\frac{1}{2n+1}\right)\frac{v_1e^3EB\bar{d}}{4\pi^2}\nonumber\\
	&\sim& j_z^{an}\ln n_{max}, \label{jz}
	\end{eqnarray}
where the right side of $~$ is valid for a large enough $n_{max}$, and $j_z^{an}=\frac{v_1e^3EBd}{4\pi^2}$ is the current rising from the chiral anomaly with $\bar{d}$ being the mean free path of the quasi-particles transforming from the right valley to the left valley \cite{NN}. We also assume that $1/d$ is much smaller than the distance between the two feroton points in the momentum space. Since the band width of the $n$th gapless mode is $4|v_1|/\sqrt{2n+1}$, then the value of $n_{max}$ is determined through  $\Delta \mu=eEd\approx 4|v_1|/\sqrt{2n+1}$, i.e.,
	\begin{equation}
	n_{max}=\frac{1}{2}(\frac{16v_1^2}{E^2\bar{d}^2}-1).
	\end{equation}
Therefore the transport current $j_z$ for a large enough $n_{max}$ can be reduced to,
	\begin{eqnarray}
	j_z\sim j_z^{an} \ln n_{max}=\frac{1}{\rho_{an}}E\ln n_{max}=\frac{E}{\rho_{MR}},
	\end{eqnarray}  
	with
	\begin{equation}
	\rho_{MR}=\frac{\rho_{an}}{\ln n_{max}}=\frac{\rho_{an}}{\ln(\frac{1}{2}(\frac{16v_1^2}{E^2\bar{d}^2}-1))},
	\end{equation}
	where $\rho_{an}=E/j_z^{an}=\frac{4\pi^2}{v_1e^3B\bar{d}}$ is the chiral anomaly contribution to the magnetoresistance as in the Weyl semimetal for $\mu<< \hbar \omega_c$ \cite{NN}. 
	}

This means that for a large $n_{max}$ ({extremely small $E$}), the resistivity {$\rho(E)$} is much smaller than $\rho_0$, the zero field resistivity. The magnetoresistance { $(\rho_{MR}(E)-\rho_0)/\rho_0\sim -1$.} As $E$ becomes stronger so that the chemical potential difference is raised to, e.g., $|v_1|/(2\sqrt {n_{max}+1/2})$ with  $n_{max}=14$,
$j_z\propto 0.0025|v_1|$, which is much smaller than the magnitude of the anomaly mode contribution which is $\propto -|v_1|$ and then gives a positive magnetoresistance whose magnitude is 400 times larger than the negative one from the chiral anomaly. Reducing $n_{max}$ to $13$, $j_z\propto -0.0315|v_1|$.  The negative magnetoresistance is 30 times large than that from the chiral anomaly. Finally,  when $\Delta\mu/2$ is larger than the maximum of the $n=2$ chiral mode energy ({see Fig. \ref{fig3}(d)}), 
{all the gapless chiral modes do not contribute except for the chiral anomaly induced lowest Landau level.} The negative magnetoresistance is the same as that in WSM \cite{ss,beenkker}. The sketch of the magnetoresistance of $E$ is shown in Fig. \ref{fig4}. Comparing to that for the WSM,  we see a more plentiful  magnetoresistance variation as the external electric field is applied {and varied.}

\section{quantum oscillations of  DOS}
\label{dos}

\subsection{Two different quantum oscillations}

Quantum oscillations directly probe the structure of the Fermi surface. For WSMs,  the quantum oscillation of the DOS between the surface states of a finite slab detects the Fermi arc structure on the surface of the slab \cite{qo}. For the feroton system, one can expect the quantum oscillation when Landau levels corresponding to the transverse ferotons pass through the Fermi level $\mu$ as $B$ varies, analog to that for the WSM. 
{To study the quantum oscillations of the feroton system, we consider the semiclassical analysis used in the quantum oscillation of the WSMs \cite{qo}. Perturbing the Hamiltonian (\ref{s1}) with a trivial square kinetic term}, i.e., for a pair of triply degenerate nodal points,
\begin{equation}
H^{\pm}_0=\pm v\boldsymbol{q}\cdot\boldsymbol{S}+\frac{1}{2m^*}q^2. \label{s2}
\end{equation}
For instance, the parameters in Eq. (\ref{GFB}) (or (\ref{GL})) are taken to be $v_1=v$, $v_2=0$ and $v_4=v_4'=\frac{1}{2m^*}$.
With an external magnetic field in the $z$-direction and taking the second term as perturbation, the spectra are given by
\begin{eqnarray}
&&E^\pm_0\approx\pm vq_z+a\hbar\omega_c^*,\nonumber\\
&&E^\pm_{1,\pm 1}\approx \pm {\rm sign}(v)\sqrt{3+q_z^2}+(1+a)\hbar\omega_c^*,\label{high}
\end{eqnarray}
 and for $n>1$,
 \begin{eqnarray}
&&E^\pm_{n,\pm1}\approx \pm {\rm sign}(v)\sqrt{2n+1+q_z^2}+(n+a)\hbar\omega_c^*,\nonumber\\
&&E^\pm_{n,0}\approx\mp\frac{v}{2n+1}q_z+(n+b)\hbar\omega_c^* \label{lower}
, \end{eqnarray}
where $a,b$ are constants about 1/2 and $\omega_c^*=\frac{eB}{m^*c}$. Taking the same geometry as that in \cite{qo}, i.e., a slab with a finite thickness $d$ in the $z$-direction and applying an external magnetic field also in the $z$-direction,  there are two periods of quantum oscillations. One of them is the same as that in WSM \cite{qo}, the WSM-like quantum oscillation which happens at a finite chemical potential. Another one presents when $\mu\approx 0$. When the bulk Fermi surface is close to the feroton nodes, the low energy states are the $E^\pm_0$ states and the surface Fermi arc states.  The closed orbit in the real space that the lower energy quasiparticles run obey the Bohr-Sommerfeld condition \cite{qo}
\begin{eqnarray}
\oint_{C}{\bf p}\cdot d{\bf r}=2\pi(n+\gamma)
\end{eqnarray}
where we take $\hbar=c=1$. Assuming $A_r$ is the area enclosed by the orbit $C$,
\begin{eqnarray}
A_r=2\pi l_B^2(n+\gamma).
\end{eqnarray}
With the magnetic length $l_B$, $Al_B^4=A_r$ defines a corresponding area in Brillouin zone, namely 
\begin{eqnarray}
A=2\pi l_B^{-2}(n+\gamma),
\end{eqnarray}
where $A$ is the area of the section of the  constant energy $E$ surface.  If $T$ is the time period that the low-lying quasiparticles circle the loop $C$, $TdE=l_B^2dA$. Integral over two sides, one has
\begin{eqnarray}
ET=2\pi l_B^2( n+\gamma).
\end{eqnarray}
The low-lying quasiparticle cycle $C$ in the present case are the same as that in the quantum oscillation of the WSM \cite{qo}.
{The period $T$ is composed of two parts. One is the time that the quasiparticle needs for running over the Fermi arcs $T_{arc}=\frac{2k_0l_B^2}{v}$ where $k_0$ is the length of the Fermi arc, and the other one is the time for the quasiparticle going through the bulk, $T_{bulk}= \frac{2d}{v}$.}

{For a finite chemical potential $\mu$, the peak of DOS of quasiparticle with energy $E_{n,1}^+$ appears when $E_{n,1}^+=\mu$, namely,}
\begin{eqnarray}
\mu&=&E_{n,1}^+=\frac{\pi v(n+\gamma_+)}{d+k_0l_B^2},\\
\Rightarrow \frac{1}{B^+_{n,1}}&=&\frac{e}{k_0}(\frac{\pi v}\mu(n+\gamma_+)-d).
\end{eqnarray}
This $1/B^+_{n,1}$ gives the peak of the quantum oscillation and the width of the interval, i.e., the period of the oscillation, is given by
\begin{eqnarray}
\Omega_{1/B_1}=\Delta\frac{1}{B_{n,1}^+}\approx \frac{d1/B_{n,1}^+}{dn}=\frac{e\pi v}{\mu k_0}.
\end{eqnarray} 
In general, $k_0\lesssim k_W$, {where $k_W$ is} the separation between two Weyl points. The chemical potential $\mu$ is taken to be a finite value. 

\begin{figure}[ptb]
\centering
\vspace{5mm}
\includegraphics[width=7cm]{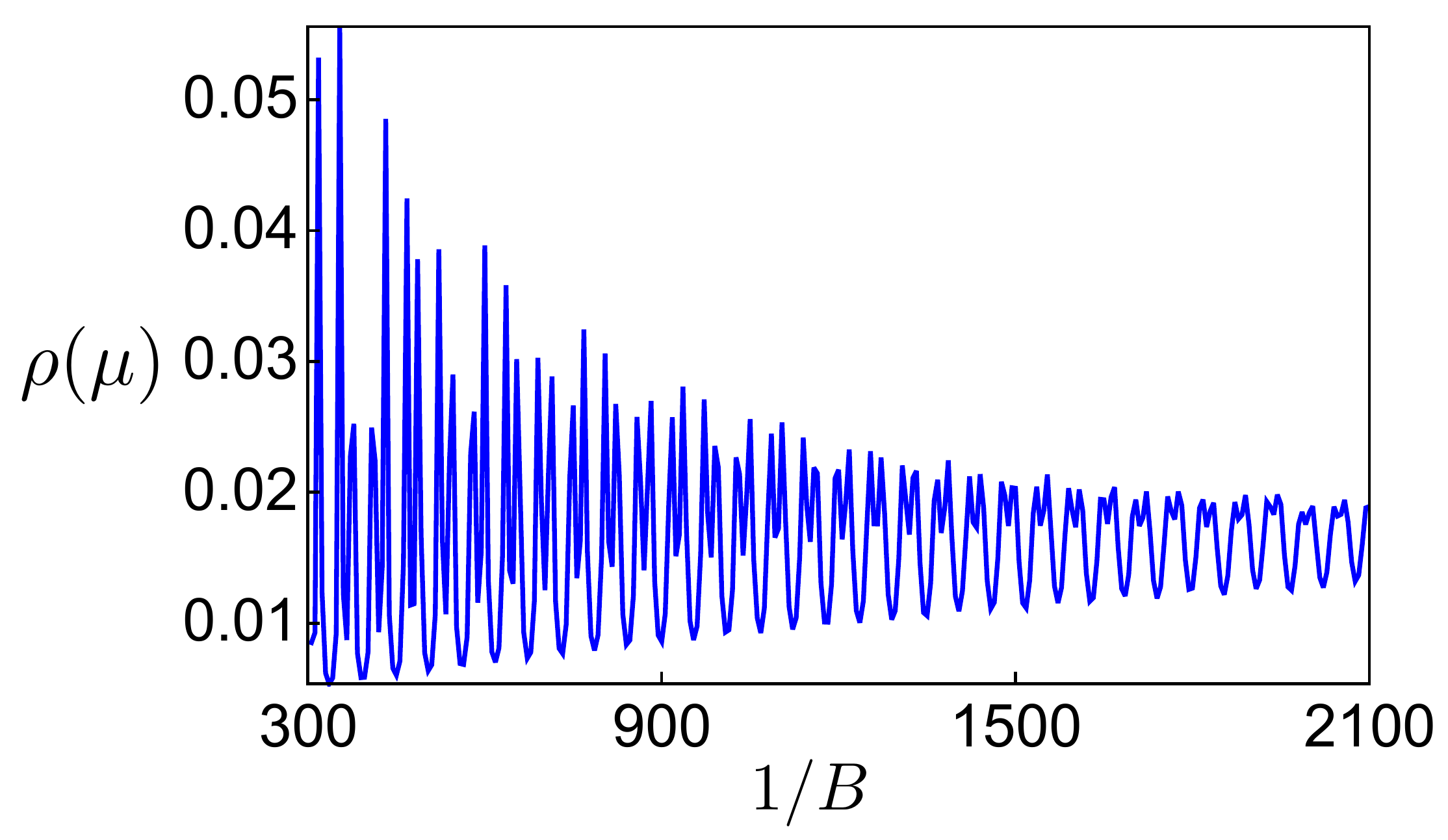}
\newline\caption{The WSM-like quantum oscillation of the density of states $\rho(\mu)$ as $1/B$ for the lattice model (\ref{GL}), where $\mu=-0.4$ and we take $m^*=10$.}
\label{fig5}
\end{figure}

This semiclassical study for the quantum oscillation of the DOS can be verified with a numerical calculation. We use the lattice model (\ref{GL}) numerically confirm this quantum oscillation of  the DOS $\rho(\mu)$. Taking Landau gauge, we use the
recursive Green's function method \cite{rg} to recursively treat the real space degrees of freedom in $\hat{x}$ direction. {See Appendix \ref{app2} for the detailed description of the numerical method.} We set system size $L_x = 10^4$ and choose a small imaginary part $\delta=10^{-3}$ for the level broadening. The thickness of the slab is 20 layers in the $z$ direction.  In Fig. \ref{fig5},  the period of the quantum oscillation of $\rho(\mu)$ is roughly a constant $\Omega_{1/B}\approx 60$ when $\mu=-0.4$ and $m^*=10$.

There is an extra quantum oscillation corresponding to $E_{n,0}^\pm$ bands. Unlike the period superposition in metal, this extra quantum oscillation happens at {a very small $\mu$, a} different energy scale from that of the WSM-like quantum oscillation. The period of this extra quantum oscillation  is estimated as follows: For the quasiparticle with energy $E^\pm_{n,0}$, defining $v^*$ by $\frac{1}2m^*v^{*2}=(n+b)\omega_c^*=(n+b)eB/m^*$ , i.e.,
$v^*=\sqrt{2(n+b)}/m^*l_B\ll v$, one has, in the semiclassical limit,
\begin{eqnarray}
E^\pm_{n,0}=\frac{\pi v(n+\gamma_{0,\pm})}{d+k_0l_B^2v/v^*}=\frac{\sqrt{2(n+b)}\pi v(n+\gamma_{0,\pm})}{\sqrt{2(n+b)}d+k_0p^*l_B^3},
\end{eqnarray}
where $p^*=m^*v$ and $\gamma_{0,\pm}$ are {constants of order unit.} Similarly, there is a peak of the DOS when $E^\pm_{n',0}=\mu$, namely,
\begin{eqnarray}
 \frac{1}{B^+_{n,0}}=\frac{e(2(n+b))^{1/3}}{(k_0p^*/d)^{2/3}}\left[\frac{\pi v}{\mu d}(n+\gamma_{0,+})-1\right]^{2/3}.
\end{eqnarray}
This $1/B^+_{n,0}$ gives the position of the peak of oscillated DOS and the width of the interval for a large $n$ is
\begin{eqnarray}
 \Omega_{1/B_0}=\Delta\frac{1}{B^+_{n,0}}\approx \frac{d}{dn}\frac{1}{B^+_{n,0}}\approx e\left(\frac{\sqrt2\pi }{\mu k_0m^*}\right)^{2/3}.\label{scp}
 \end{eqnarray}
Notice that since the band widths of $E_{n,0}$ are very narrow, the Fermi level must be as close to the triply degenerate nodal point as possible, i.e, $\mu\approx0$ in order to see the extra quantum oscillation. On the other hand, since  the Fermi level is very close to the triply degenerate {feroton} point, the length of the measured Fermi arc $k_0$ may be closer to $k_W$ than that in the WSM-like quantum oscillation.

\begin{figure}[ptb]
\centering
\vspace{5mm}
\includegraphics[width=7cm]{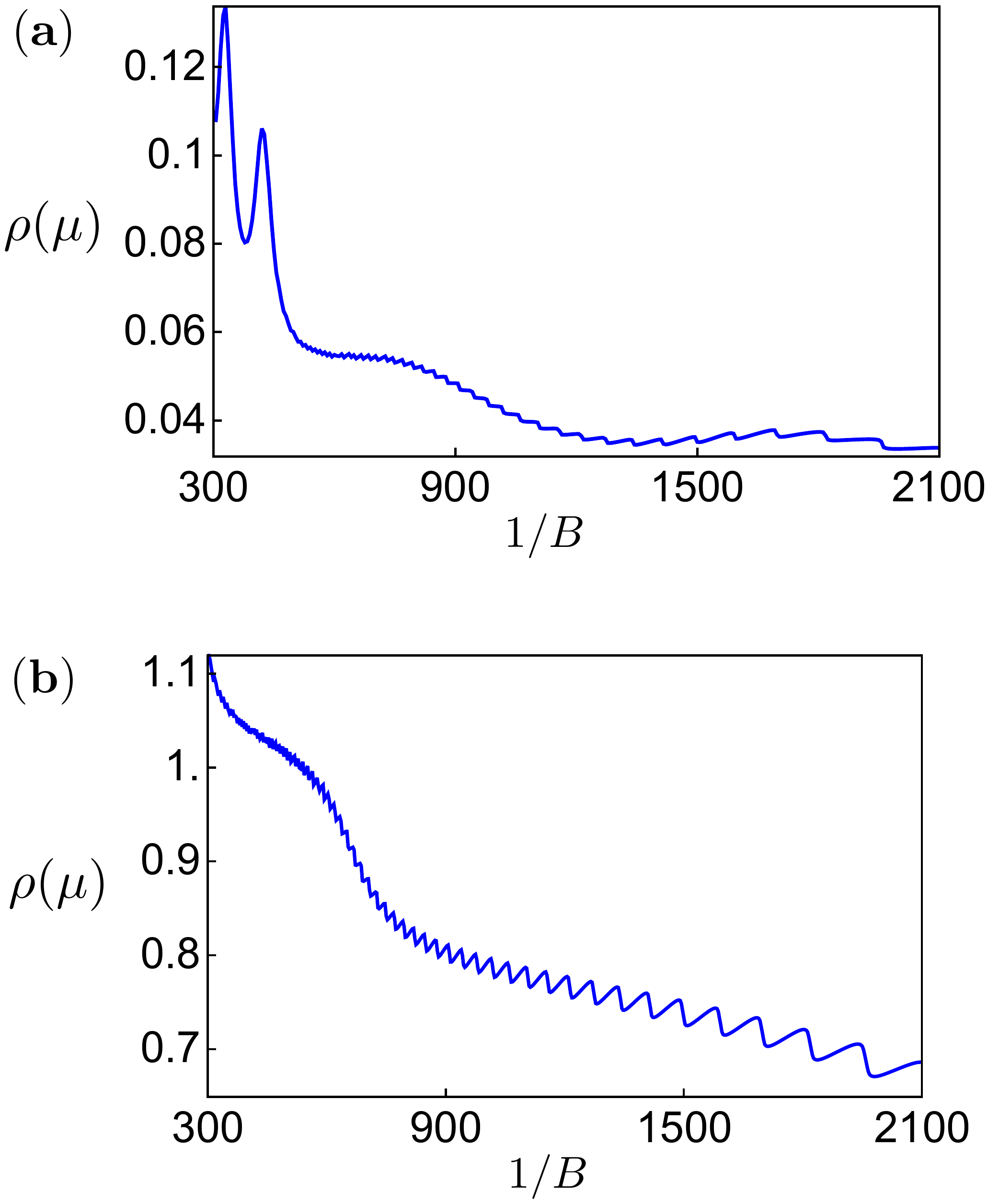}
\newline\caption{The extra quantum oscillations of the  density of states for the lattice model (\ref{GL}). We take $\mu=0$ and (a) $m^*=2$; (b) $m^*=10$.}
\label{fig6}
\end{figure}

We numerically study this extra quantum oscillation at $\mu\approx 0$ with the same parameters as those before. Fig. \ref{fig6} shows that  both the amplitude and the period of the oscillation are very sensitive to $m^*$. The flatter the quadratic term is, the larger the amplitude is, as expected. On the other hand, the period is shorter as $m^*$ becomes larger, which is qualitatively consistent with the semiclassical result (\ref{scp}).  The result for $m^*=10$ shows that the period of the oscillation is gently dependent on $1/B$, which is slightly different from (\ref{scp}).

 \subsection{Quantum oscillation without Landau gap}

 To further explore the effects brought by these gapless modes, we study the following model on a cubic lattice with
  \begin{eqnarray}
    H&=&\sum_i t c_i^\dagger S^1 c_{i\pm\hat{x}} \mp it c_i^\dagger S^2 c_{i\pm\hat{y}} \mp it c_i^\dagger S^3 c_{i\pm\hat{z}} \label{lh} \\
    && -\frac{\Delta}{2}\left(c_i^\dagger S^1 c_{i\pm\hat{y}}+c_i^\dagger S^1 c_{i\pm\hat{z}}\right)+2\left(\Delta-t\cos k_0\right) c_i^\dagger S^1 c_i \nonumber \\
    && -\frac{1}{2m^*}\left(c_i^\dagger c_{i\pm\hat{y}} +
    c_i^\dagger c_{i\pm\hat{z}}\right),\nonumber
    \end{eqnarray}
    where $c_i$ is a three component fermion operator, i.e, there are three flavors of fermions at each site. Taking  proper parameters, say, $t=1$, $K=\frac{\pi}{2}$, $\Delta=4$, there are two triply degenerate nodal points $(\pm K,0,0)$. The spectra near the nodal points are given by
   \begin{eqnarray}
    E_0({\bf k})&=& \frac{1}{2m^*}(k_y^2+k_z^2), \nonumber \\
    E_\pm({\bf k})&=& \pm\sqrt{(k_x\pm K)^2+k_y^2+k_z^2}+E_0({\bf k}).
    \end{eqnarray}
Different from the spectra (\ref{s2}), there is no quadratic term of $k_x\mp K$ in $E_0$.  When $m^*\geq1$, there are no other nodal points at $E=0$ besides the triple degenerate nodal points. For $m^*\to \infty$, this is the transverse feroton model. The transverse ferotons give the WSM-like quantum oscillation of the DOS. For a small quadratic term, the qualitative behavior of the WSM-like quantum oscillation is the same as that studied in the last subsection.

To see the extra quantum oscillation, we apply the magnetic field perpendicular to the $x$-$y$ plane.   Taking Landau gauge $\mathbf{A}=(0,-Bx,0)$, $k_y$ is a good quantum number which gives the position $x_0$ of the peak of a Landau orbit, i.e., $k_y=x_0/l_B^2=x_0B$ with $0<x_0<L_x$. Thus, besides the finite energy Landau energy, there is an energy band which consists of the restored gauge degrees of freedom and $E_{0,0}(k)$. 

\vspace{2mm}

\centerline{\bf B1. Continuous limit} 

\vspace{2mm}

We consider a slab which is thick enough so that $k_z$ can be thought as a good quantum number. In the long wavelength limit,  the spectra read
\begin{eqnarray}
E_{0,0}(k_y,k_z) &=&-k_z+\frac{1}{2m^*}(k_y^2+k^2_z),\nonumber\\
E_{n,0}(k_y,k_z)&=&\frac{k_z}{2n+1}+\frac{1}{2m^*}(k_y^2+k^2_z),~n>1.
\end{eqnarray}
The DOS for $E_{n,0}$ is given by

\begin{equation}
\rho_{n,0}(E)=\left|\frac{1}{\frac{\partial E_{n,0}}{\partial {k_z}}}\right|=1/\left|\left(\frac{1}{2n+1}+\frac{k_z}{m^*}\right)\right|.
\end{equation}
Since $k_z\sim (2n+1)(E_{n,0}-\frac{k_y^2}{2m^*})$, then,
\begin{eqnarray}
\rho_{n,0}&=&1/\left|\left(\frac{1}{2n+1}+\frac{2n+1}{m^*}\left(E_{n,0}-\frac{x^2_0B^2}{2m^*}\right)\right)\right|\nonumber\\
&=&\frac{2n+1}{\left|1+\frac{(2n+1)^2}{m^*}\left(E_{n,0}-\frac{x_0^2B^2}{2m^*}\right)\right|}.\label{ds}
\end{eqnarray}

 At the half-filing, taking $E_{n,0}=\mu=0$, the DOS on the Fermi surface is given by
\begin{eqnarray}
\rho_{n,0}(0)\approx\frac{2m^{*2}(2n+1)}{\left|2m^{*2}-(2n+1)^2x_0^2B^2\right|},
\end{eqnarray}
which is singular at $\frac1{B_{n,0}}=\frac{(2n+1)x_0}{\sqrt2 m^*}$ and the period of the quantum oscillation  is given by
\begin{eqnarray}
\Omega_{1/B_0}=\Delta\frac{1}{B_{n,0}}=\frac{\sqrt2x_0}{m^*}. \label{m}
\end{eqnarray}
In reality, the singularity in (\ref{ds}) is always destroyed by the level broadening $\delta$ and the peak of the DOS is proportional to $2m^{*2}(2n+1)$.

\vspace{2mm}

\centerline{\bf B2. Numerical calculation for a finite thickness slab}

\vspace{2mm}

We numerically demonstrate this low energy extra quantum oscillation  for the lattice model (\ref{lh}). 
$L_x$ is the size in the $x$ direction and the thickness of the slab is 40 layers in the $z$ direction. Since $k_z$ is no longer a good quantum number in the slab,  the quantitative matching between the {numerical} calculations with this finite thickness slab and the {results in the} continuous limit would not be expected in {the numerical} calculation. What we want to show {through the numerical calculations} is the existence of the quantum oscillation in such a {finite slab geometry}, which may be experimentally relevant. We use the recursive Green's function method \cite{rg} (for details, see Appendix A). As shown in Fig. \ref{fig7}, $\rho(0)$ is oscillated with varying of $B$.  We see that the amplitudes of the quantum oscillation of the DOS $\rho$ increases as $m^*$ increasing while the period is weakly dependent on $m^*$. As expected before, this is not quantitatively matched with (\ref{m}).   Numerical result also shows the quantum oscillation is not dependent on the choice of $x_0$, which is not quantitatively consistent with (\ref{m}).  
Beside of the finite thickness of the slab, another reason for this mismatching is that the open boundary
condition in the continuous limit is implied while the periodic boundary condition on the $x-y$ plane is taken in
the lattice model study.

\begin{figure}[ptb]
	\centering
	\vspace{5mm}
	\includegraphics[width=7cm]{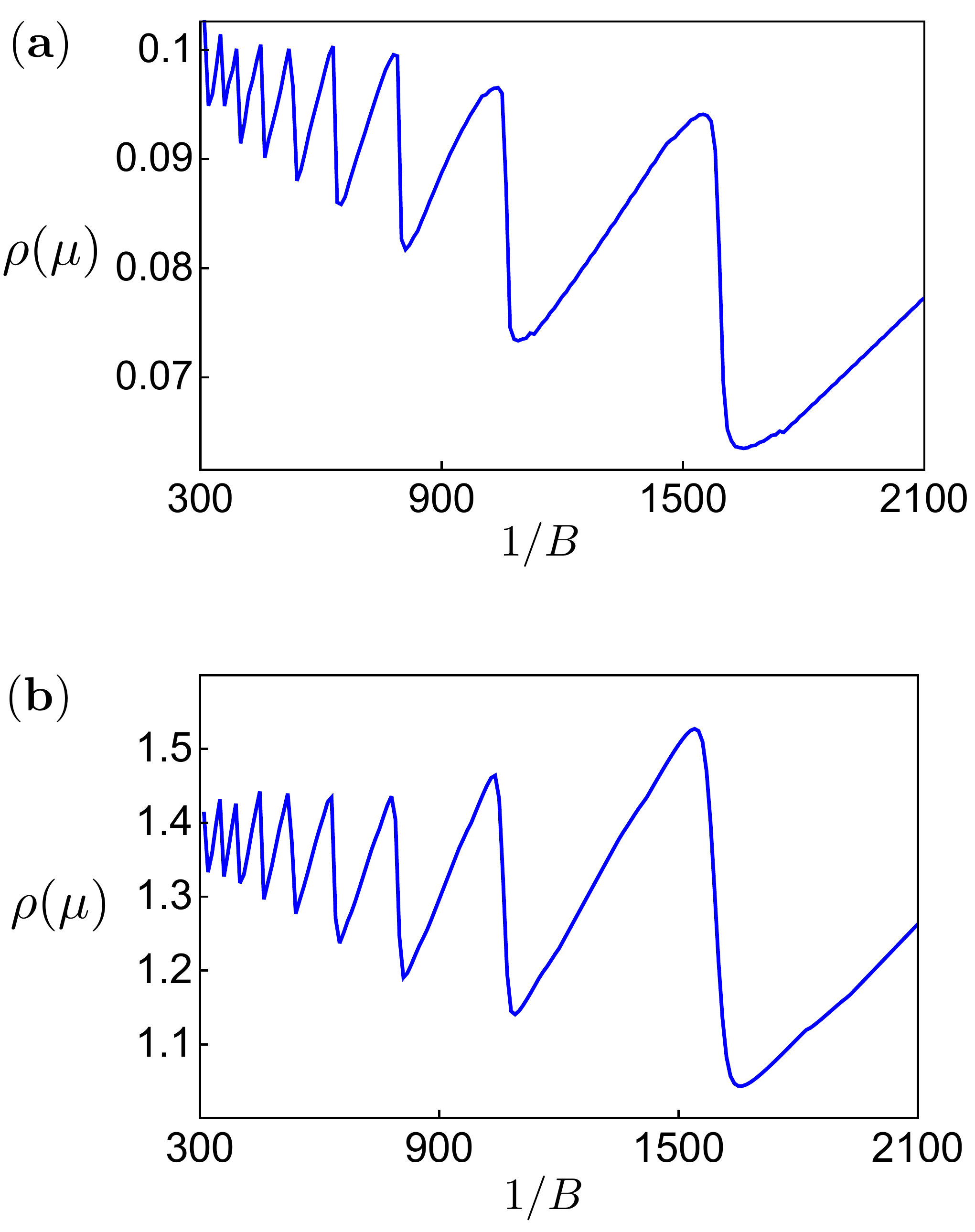}
	\newline\caption{The extra quantum oscillation of the density of states for the lattice model (\ref{lh}). We take $\mu=0$ and (a) $m^*=2$; (b) $m^*=20$.}
	\label{fig7}
\end{figure}

\section{possible materials with feroton excitations} 

We have studied interesting spectroscopic and magnetic transport properties of the charged feroton excitations in condensed 
system. The evidences of the feroton excitations already exhibited in several condensed matter systems.  In Ref. \cite{ber}, it was predicted that a feroton dispersion relation emerges at a high symmetry point in Brillouin zone of a body-centered cubic Bravais
lattice with space groups 199 and 214.  For examples, Pd$_3$Bi$_2$S$_2$ with space group 199 has a triply degenerate node at $P$ point and almost at the Fermi level and Ag$_3$Se$_2$Au with space grou 214 has such a node 0.5 eV below the Fermi level \cite{ber}. 

 In simple half-Heusler topological insulators, the triply degenerate nodal points were also predicted \cite{tdnp1}. 
A feroton spectrum with quadratic perturbation was found for LaPtBi exactly at $\Gamma$ point. 
{This provides a promising candidate for testing our predictions} of the magnetic transport properties of the ferotons.  

In a recent {\it ab initio} calculation, the triply degenerate nodel points are also found in materials with symmorphic structures \cite{tdnp4}.  Among these materials, the band structure of ZrTe owns feroton dispersions at two triply degenerate nodal points both of which are very close to the Femi level.

 \section{ Conclusions.}

  We have studied 
  {some novel physical properties of}
  the photon-like fermionic semimetal, the feroton semimetal.  We pointed out that the exact flat bands at the Fermi level in a band theory are the redundant degrees of freedom, similar to the longitudinal photon in Maxwell theory. The external magnetic field breaks the fermionic gauge symmetry  and a bunch of  gapless elementary excitations 
  {emerge from the longitudinal feroton. Each of the gapless modes carries a different Landau index $n>1$ which provides the orthogonality of the gapless modes.} Due to the existence of these gapless excitations, the magnetic transport phenomena of the system become very plentiful. Interesting extra low energy quantum oscillations of the DOS are shown both with or without the low energy Landau gaps. 
  {The feroton may emerge at high symmetry points of space group 199, 214, and 220 \cite{ber}, for instance, Pd$_3$Bi$_2$S$_2$, Ag$_3$Se$_2$Au \cite{ber}, LPtBi(L=La,Yp,Gd) \cite{tdnp1} and ZrTe \cite{tdnp4}. Therefore we expect these novel} phenomena are experimentally observable.

\noindent{\bf Acknowledgments}
We thank Gang Chen and Fei Teng for helpful discussions. This work was supported by NNSF of China (Nos. 11774066 and11474061, XL,FYL,YY) and the ministry of science and technology of China with the grant No.2016YFA0301001 (FYL).

\appendix

\section{The magnetoresistance of feroton semimetal}\label{app1}

In this appendix, we present a detailed analysis on the magnetoresistance of feroton semimetal through the Boltzmann equation in the quantum limit \cite{NN}. For a strong magnetic field B, the chemical potential $\mu<<\hbar \omega_c$, where $\omega_c$ is the cyclotron frequency. In the quantum limit, only the gapless modes will participate in the transport. As discussed in the main text, each of the gapless modes is associated with a different quantum number, the Landau level index $n$, thus the gapless modes are orthogonal to each other. In the following we assume that the scattering between different Landau levels is almost zero. Following a similar treatment in Ref. \cite{NN}, for a given gapless mode with Landau level index $n$, its low energy spectrum reads,
\begin{equation}
E_n^0\sim V_nq_z,
\end{equation} 
where $V_n$ is the slope of the spectrum near $q_z=0$, especially, if we denote $V_0=v_1$ for the Lowest Landau level, then for $n>1$, $V_n=-\frac{v_1}{2n+1}$. For simplicity, we assume there is only one pair of feroton points in the system, which we denote as L and R. The density per length l is $elB/4\pi^2$, then after applying an electric field E parallel to B, the states move along E according to $\dot{q_z}=eE$. Therefore the change rate of quasi-particle number $N_R^n$ in a given gapless band $n$ near the R valley reads,
\begin{equation}
\dot{N_R^n}=\frac{1}{l}\frac{leB}{4\pi^2}\dot{E_0^n}=V_n\frac{e^2}{4\pi^2}EB.
\end{equation}
Similarly, the creation rate of anti quasi-particle near the L valley reads,
\begin{equation}
\dot{\bar{N}}_L^n=V_n\frac{e^2}{4\pi^2}EB.
\end{equation}
Therefore the drift rate reads,
\begin{equation}
\dot{N}_{drift}=V_n\frac{e^2}{4\pi^2}EB.
\end{equation}
Now let us consider the scattering effects in the system. The scattering within one valley is neglectable \cite{NN}, and we denote the relaxation time $\tau$ for the scattering between the L and R valleys. Then the collision term of one valley, say the R valley, reads,
\begin{equation}
\dot{N}_R^n|_{coll}=-\frac{N_R^n-N_{R0}^n}{\tau},
\end{equation} 
where $N_{R0}^n$ denotes the quasi-particle number in the R valley when $B=0$. Since at the zero temperature, the Fermi-Dirac distribution is $f_0(\epsilon)=\theta(\mu_R-\epsilon)$ at the R valley, then,
\begin{equation}
N_R^n=\frac{1}{l^3}\sum_{q_y,q_z}f_0(\epsilon)=\frac{eB \mu_R}{4\pi^2V_n}.
\end{equation}
The results of the L valley are also similar to those of the R valley. Therefore, using the Boltzmann equation,
\begin{eqnarray}
\dot{N}_R^n|_{drift}=-\dot{N}_R^n|_{coll},
\end{eqnarray}
we have,
\begin{equation}
\Delta\mu=\mu_R-\mu_L=eEd_n,
\end{equation}
where $d_n=V_n \tau$ being the mean free path. For simplicity, we further assume that $d_n$ are the same for all $n$ which we denote as $\bar{d}$, and $1/d$ is much smaller than the distance between two feroton points in the reciprocal space. 

Since the transfer rate of quasi-particles from R valley to L valley is $V_n\frac{e^2EB}{4\pi^2}$ per unit time and unit volume, the total energy cost $V_n\frac{e^2EB}{4\pi^2}\Delta \mu$ is provided by the external electric field E, namely,
\begin{equation}
EJ_n=V_n\frac{e^2EB}{4\pi^2}\Delta \mu=V_n\frac{e^2EB}{4\pi^2}eE\bar{d},
\end{equation}
thus the current $J_n$ contributed from a given gapless band $n$ reads,
\begin{equation}
J_n=\frac{e^3V_nEB\bar{d}}{4\pi^2}.
\end{equation}

Notice that the number of bands that participate in the transport is related to $\Delta \mu$. Because the band width of the nth gapless mode is $4|v_1|/\sqrt{2n+1}$, then $n_{max}$ is determined through $\frac{4|v_1|}{\sqrt{2n_{max}+1}}\approx\Delta \mu$, i.e, 
\begin{equation}
n_{max}=\frac{1}{2}((\frac{4|v_1|}{eE\bar{d}})^2-1).
\end{equation} 
Therefore, only the $n<n_{max}$ gapless bands contribute to the transport. Therefore the total current $J$ reads,
\begin{equation}
J=\sigma E=\sum_{n=0}^{n_{max}}J_n=(\sum_{n=0}^{n_{max}}V_n)\frac{e^3EB\bar{d}}{4\pi^2}.
\end{equation}
For a large enough $n_{max}$, the conductivity $\sigma$ reads
\begin{equation}
\sigma=\frac{v_1e^3B\bar{d}}{4\pi^2}\ln n_{max},
\end{equation}
and the magnetoresistance reads,
\begin{equation}
\rho_{MR}=\frac{1}{\sigma}=\frac{1}{\ln n_{max}}\frac{4\pi^2}{v_1e^3Bd}=\frac{\rho_{MR}^{an}}{\ln (\frac{1}{2}((\frac{4|v_1|}{eEd})^2-1))},
\end{equation}
with $\rho_{MR}^{an}=\frac{4\pi^2}{v_1e^3Bd}$ being the contribution of the Lowest Landau level for $\mu<<\hbar \omega_c$ which is related to the chiral anomaly \cite{NN}.

In the regime $\mu>>\hbar \omega_c$, all the gapless modes but the chiral anomaly one quit transport. Then the magnetoresistance reduces to the well-known result of the Weyl semimetal \cite{ss,beenkker},
\begin{equation}
\rho_{MR}=\frac{2\pi^2\mu^2}{e^3v_1^3B^2\tau'},
\end{equation}
where $\tau'$ is an elastic inter-valley scattering mean free
time \cite{ss}. This magnetoresistance is half smaller than that in the Weyl semimetal because the monopole number of feroton point is two. For a large chemical potential, the chiral anomaly comes from the $E_1^+$ band and the chiral lowest Landau level. Each of these two bands has a band Chern number one. Therefore the total current will be twice as big as that in the Weyl semimetal \cite{ss}. 

\section{The recursive Green's function method} \label{app2}
In the calculation of the density of states (DOS) in Sec.\ref{dos}, we use the recursive Green's function method described in Ref.~\onlinecite{rg,rg1,rg2}.
For completeness, here we summarize the formulae to calculate the DOS.

For a quadratic tight-binding Hamiltonian $\mathcal{H}=\sum_{ij}
H_{ij} |i\rangle \langle j|$, we can define the single particle Green's function $\mathcal{G}^{\pm}(z)$ as
\begin{equation}
(z^{\pm}\mathcal{I}-\mathcal{H})\mathcal{G}^{\pm}=\mathcal{I},
\end{equation}
where $\mathcal{I}$ is the identity operator, $z^{\pm}=\mu\pm i\delta$ is a complex variable and a small imaginary part $\delta$ is chosen for energy level broadening in practical calculation.

The advanced and the retarded Green's functions are defined as $\mathcal{G}^-(\mu-i0)$ and
$\mathcal{G}^+(\mu+i0)$, respectively. In the basis of the functions $|i\rangle$,
$\mathcal{G}^{\pm}=\sum_{ij} G^{\pm}_{ij} |i\rangle \langle j|$ and we have
\begin{equation}
(z^{\pm}\delta_{ij}-H_{ij})G^{\pm}_{jk}=\delta_{ik}.
\end{equation}
Note that $G^-_{ij}=(G^+_{ji})^*$ due to the hermiticity of $H$.

If the Hamiltonian can be arranged in a block matrix form, containing only 
the ``nearest-neighbor hopping'' matrix elements,
\begin{equation*}
H=
\begin{pmatrix}
\ddots & & & \\
\ldots & \textbf{H}_{ii} & \textbf{H}_{i,i+1} & \ldots \\
\ldots & \textbf{H}_{i+1,i} & \textbf{H}_{i+1,i+1} & \ldots \\
& & & \ddots
\end{pmatrix},
\end{equation*}
where \textbf{H} is a block matrix, we then get
\begin{equation}
(z\textbf{I}-\textbf{H}_{ii}) \textbf{G}_{ij}
-\textbf{H}_{i,i-1}\textbf{G}_{i-1,j}-\textbf{H}_{i,i+1}\textbf{G}_{i+1,j}
=\textbf{I}\delta_{ij}.
\end{equation}
Note we omit the superscript $\pm$ in $z$ and $\textbf{G}$.

We can construct a larger system by adding 
a new slice to the system consisting of $N$ slices: Adding blocks $\textbf{t}_N$, $\textbf{t}^\dagger_N$ 
and $\textbf{H}_{N+1,N+1}$ to $H^{(N)}$, we obtain $H^{(N+1)}$. 
We use the notation $\textbf{t}_i\equiv\textbf{H}_{i,i+1}$ and $\textbf{t}_i^\dagger\equiv\textbf{H}_{i+1,i}$. 

More important, we can calculate the Green's function of the larger system recursively using the following relations:
\begin{align}
&\textbf{G}^{(N+1)}_{N+1,N+1}=\left[z\textbf{I}-\textbf{H}_{N+1,N+1}-
\textbf{t}_N^\dagger \textbf{G}^{(N)}_{NN}\textbf{t}_N\right]^{-1}, \\
&\textbf{G}^{(N+1)}_{ij}=\textbf{G}^{(N)}_{ij}+
\textbf{G}^{(N)}_{i,N}\textbf{t}_N\textbf{G}^{(N+1)}_{N+1,N+1}
\textbf{t}_N^\dagger \textbf{G}^{(N)}_{N,j} \quad (i,j \leq N), \nonumber \\ \\
&\textbf{G}^{(N+1)}_{ij}=\textbf{G}^{(N)}_{ij}+
\textbf{G}^{(N)}_{i,N}\textbf{t}_N\textbf{G}^{(N+1)}_{N+1,N+1}
\textbf{t}_N^\dagger \textbf{G}^{(N)}_{N,j} \quad (i,j \leq N), \nonumber \\ \\
&\textbf{G}^{(N+1)}_{i,N+1}=\textbf{G}^{(N)}_{i,N}
\textbf{t}_N\textbf{G}^{(N+1)}_{N+1,N+1} \quad (i \leq N), \\
&\textbf{G}^{(N+1)}_{N+1,j}=\textbf{G}^{(N+1)}_{N+1,N+1}
\textbf{t}_N^\dagger \textbf{G}^{(N)}_{N,j} \quad (j \leq N).
\end{align}

The DOS is given by
\begin{equation}
\rho(\mu)=-\frac{1}{\pi\Omega} \text{Im Tr }\mathcal{G}^+=\frac{1}{\pi N M^2}\text{Im}
\sum_{i=1}^{N}\text{Tr } \textbf{G}^+_{ii},
\end{equation}
where $\Omega$ is the volume of the system, $N$ is the number of slices and $M^2$ is the number of sites in each slice.
Introducing two auxiliary matrices
\begin{eqnarray}
&&\textbf{R}_N=\textbf{G}^+_{N,N}, \\
&&\textbf{F}_N=\textbf{t}_N^\dagger \left[ \sum_{i=1}^{N} \textbf{G}^{+(N)}_{N,i}
\textbf{G}^{+(N)}_{i,N} \right] \textbf{t}_N,
\end{eqnarray}
we have following recursive relations:
\begin{align}
&\textbf{R}_{N+1}=\left[z^+\textbf{I}-\textbf{H}_{N+1,N+1}-
\textbf{t}_N^\dagger \textbf{R}_N \textbf{t}_N\right]^{-1}, \\
&\textbf{F}_{N+1}=
\textbf{t}_{N+1}^\dagger \textbf{R}_{N+1} (\textbf{F}_N+\textbf{I})
\textbf{R}_{N+1} \textbf{t}_{N+1}, \\
& s_\rho^{(N+1)}=s_\rho^{(N)} + \text{Tr }\{\textbf{R}_{N+1}(\textbf{F}_N+\textbf{I})\}.
\end{align} 
Note that the initial values can be deduced from the recursive relations, consistent with their definitions:
\begin{eqnarray}
&&\textbf{R}_{1}=\left[z^+\textbf{I}-\textbf{H}_{1,1}\right]^{-1}, \\
&&\textbf{F}_{1}=
\textbf{t}_{1}^\dagger \textbf{R}_{1} 
\textbf{R}_{1} \textbf{t}_{1}, \\
&& s_\rho^{(1)}=\text{Tr } \textbf{R}_{1}.
\end{eqnarray}
Finally we have
\begin{equation}
\rho^{(N+1)}(\mu)=-\frac{1}{\pi (N+1) M^2} \text{Im }s_\rho^{(N+1)}.
\end{equation}

For our model Hamiltonians, the thickness in $\hat{z}$ direction $L_z$ is taken as finite and the magnetic field is applied as $\mathbf{B} = B\hat{z}$. In Landau gauge $\mathbf{A} = (0, -Bx, 0)$, $k_y$ is still a good quantum number and the DOS is independent of the choice of $k_y$ in the thermodynamic limit, i.e. for large enough system size in $\hat{x}$ direction. The real space degrees of freedom in $\hat{x}$ direction are then treated recursively in numerical calculation and the number of iterations determines the system size in $\hat{x}$ direction $L_x$. In practical calculation, one need set a small $\delta$ for energy level broadening. The numerical accuracy gets better with larger $L_x$ and smaller $\delta$. We typically choose $L_x=10^4$ and $\delta=10^{-3}$ in our calculation, since further optimization doesn't change the numerical results obviously.

\end{document}